\documentclass[prx,twocolumn,floatfix,superscriptaddress]{revtex4-2}
\usepackage{graphicx}
\usepackage[utf8]{inputenc}
\usepackage{amssymb}
\usepackage{amsmath}
\usepackage[colorlinks,allcolors=blue]{hyperref}
\usepackage{braket}
\usepackage{placeins}
\usepackage{soul}
\usepackage{etoolbox}

\begin{document}

\setlength\abovedisplayskip{5pt}
\setlength\belowdisplayskip{5pt}

\title{Timescales of quantum and classical chaotic spin models evolving toward equilibrium}
\author{Fausto Borgonovi}
\affiliation{Dipartimento di Matematica e
  Fisica and Interdisciplinary Laboratories for Advanced Materials Physics,
  Universit\`a Cattolica, via della Garzetta  48, 25133 Brescia, Italy}
\affiliation{Istituto Nazionale di Fisica Nucleare,  Sezione di Milano,
  via Celoria 16, I-20133,  Milano, Italy}
\author{Felix M. Izrailev}
\affiliation{Instituto de F\'{i}sica, Benem\'{e}rita Universidad Aut\'{o}noma
  de Puebla, Apartado Postal J-48, Puebla 72570, Mexico}
\affiliation{Department of Physics and Astronomy, Michigan State University, E. Lansing, Michigan 48824-1321, USA}
\author{Lea F. Santos}
\affiliation{Department of Physics, University of Connecticut, Storrs, Connecticut 06269, USA}


\begin{abstract}
We investigate quench dynamics in a one-dimensional spin model, comparing both quantum and classical descriptions. Our primary focus is on the different timescales involved in the evolution of the observables as they approach statistical relaxation. Numerical simulations, supported by semi-analytical analysis, reveal that the relaxation of single-particle energies (global quantity) and on-site magnetization (local observable) occurs on a timescale independent of the system size $L$. This relaxation process is equally well-described by classical equations of motion and quantum solutions, demonstrating excellent quantum-classical correspondence, provided the system be strongly chaotic. The correspondence persists even for small quantum spin values ($S=1$), where a semi-classical approximation is not applicable. Conversely, for the participation ratio, which characterizes the initial state spread in the many-body Hilbert space and which lacks a classical analogue, the relaxation timescale is system-size dependent. 
\end{abstract}

\maketitle

\section{Introduction}
The relaxation process of isolated many-body quantum systems toward statistical equilibrium, and the dependence of their relaxation times on model parameters, has recently attracted significant attention. However, different studies have reported conflicting conclusions regarding how the relaxation time scales with system size $L$. Some investigations suggest that the relaxation time decreases with increasing $L$~\cite{Goldstein2013,Goldstein2015}, while other find it weakly on $L$~\cite{Oliveira2018,Carvalho2023}, or entirely independent of $L$~\cite{Niknam2021}. Several studies also indicate that the relaxation time increases  with $L$ \cite{Reimann2008,Reimann2016,Short2011,Short2012,Monnai2013,Goldstein2013,Malabarba2014,Hetterich2015,Gogolin2016,Pintos2017,Schiulaz2019,Bertini2021,Dymarsky2022} either polynomially or exponentially, depending on the observable~\cite{Lezama2021}. The ongoing debate persists due to the absence of a general theoretical framework and the intrinsic challenges associated with simulating quantum systems with a large number of particles.

Recently, we proposed a novel approach~\cite{Benet2023} to describe the statistical properties of interacting many-body quantum systems with a well-defined classical limit. The main idea is that, under the condition of strong chaos in both quantum and classical models, certain global characteristics of the eigenstates can be derived from the classical equations of motion. Since the theoretical analysis of classical systems is relatively simpler than that of quantum systems, this approach allows for semi-analytical results to be obtained based on classical chaos properties, which can then be effectively applied to the quantum systems. In Ref.~\cite{Benet2023}, we demonstrated the effectiveness of this approach using a one-dimensional (1D) spin model with varying short-range interactions. Through detailed numerical analyses, we showed that quantities that serve as building blocks of physical observables coincide in the classical and quantum descriptions.
One such quantity is the local density of states (LDoS).

The LDoS determines the energy distribution of the initial state in quench dynamics. In nuclear physics, it is known as strength function and is used to study the scattering properties of particles in nuclear reactions~\cite{ZelevinskyRep1996}.  The width of the quantum LDoS characterizes the growth rate of the participation ratio, which gives the number of many-body states participating in the evolution of an initially excited state~\cite{Borgonovi2019,Borgonovi2019b}. Additionally, the absolute square of the Fourier transform of the LDoS is the survival probability of the initial state. As detailed in~\cite{Benet2023}, for systems with a well-defined classical limit, the LDoS can be obtained from the classical trajectories associated with the non-interacting Hamiltonian $H_0$ by projecting them onto the total Hamiltonian $H = H_0+V$, where $V$ represents the inter-particle interaction. The classical and quantum LDoS coincide when the system is strongly chaotic, even for small quantum spin values $S$.

The goal of the present paper is to investigate the quantum-classical correspondence (QCC) for strongly chaotic many-body systems out of equilibrium. This is challenging, because the phase space of classical many-body models is multidimensional and the Hilbert space of the quantum many-body models grows exponentially with the number of particles, which makes the QCC analysis nearly intractable. Recent studies in this direction have been done in the context of the out-of-time-ordered correlator~\cite{Rammensee2018, Hummel2019}, and have explored spin models~\cite{Akila2017, Akila2018, Schubert2021} and a $p$-spin glass model~\cite{Correale2023}, though many questions remain open. Our approach is inspired by the QCC framework developed for many-body spin models in~\cite{Benet2023}, which we now extend to analyze the dynamical properties of these systems. The focus is on the different timescales identified along the relaxation process of various observables. With our approach, we can use the classical model to obtain semi-analytical expressions that successfully describe the quantum evolution.

Our study concentrates on the classical and quantum evolution toward equilibrium of both global and local observables, with the goal of  estimating their relaxation time. Specifically, we consider a 1D model of $L$ interacting spins in the chaotic regime and investigate the evolution of the energy associated with the non-interacting Hamiltonian, which is a global quantity, and of the on-site magnetization, which is a local observable. Our numerical results and analytical estimates reveal excellent QCC for the evolution of both quantities. We also verify that in a time window before saturation, the energy variance increases linearly over time, which indicates a diffusive-like spreading. 

Due to strong chaos and ergodicity of the classical motion of individual spins on 3D spheres, the energy spread leads to the ergodic filling of the many-dimensional energy shell. Surprisingly, we find that the timescale for the diffusive spread of energy is independent of the number of spins. The same holds for the relaxation time of the magnetization of individual spins. The reasons for these results lie in the structure of the phase space, where each spin is constrained to the surface of its 3D unit sphere, and in the choice of  uncorrelated frequencies for the single particles.

We also investigate the evolution of the participation ratio, $P_{R}(t)$, also known as the number of principal components. In contrast to the energy and magnetization, $P_{R}(t)$ has no classical analogue. This quantity measures the number of many-body states that characterize the spread of the initial state in the Hilbert space, being thus equivalent to the exponential of a participation entropy. Our numerical data confirm that $P_{R}(t)$ exhibits exponential growth over time before reaching saturation, with a growth rate determined by the width of the LDoS. We find that saturation occurs at a time that increased with the number of spins $L$. This observation is consistent with existing results for interacting fermions and bosons~\cite{Borgonovi2016, Borgonovi2019}. Our analytical study further shows that the relaxation time is proportional to $\sqrt{L}$.  

The paper is organized as follows. In Sec.~\ref{SecModel}, we introduce the spin model in the quantum and classical domain.
In Sec.~\ref{SecEnergy}, we analyze the quantum and classical spread of the single-particles energies, identifying the timescales for ballistic and diffusive behaviors, before saturation, and compare the diffusion time with the Lyapunov time, which is a timescale characteristic of chaotic systems. In Sec.~\ref{SecNameIV}, we study the dynamics of the magnetization in the $z$-direction of individual spins, find agreement with the timescales for the energy spreading, and    explain why the relaxation time for energy and magnetization are independent of the system size. In Sec.~\ref{Sec:NPC}, we investigate the evolution of the participation ratio, which has no classical analogue, and show that its relaxation time depends on system size. Finally, conclusions are presented in Sec.~\ref{Sec:Discussions}.

\section{Quantum and Classical model} 
\label{SecModel}

We consider the same  model explored in Ref.~\cite{Benet2023}. The total Hamiltonian,

\begin{equation}
 H=H_0 +V ,
\label{eq:ham0a}
\end{equation}
consists of two parts. 
The first part, 
\begin{equation}
 H_0 = \sum_{k=1}^L B_k S_k^z ,
\label{eq:ham0a}
\end{equation}
 describes $L$ non-interacting spins on a 1D lattice in a slightly  inhomogeneous magnetic field along the $z$-axis.  $B_k$ are the local frequencies associated with each spin. We consider an almost homogeneous distribution of the single particle frequencies  $B_k = B_0 +\delta B_k$, where $B_0 = 1 $ and $\delta B_k$ are small random entries, $|\delta B_k| \leq \delta W \ll B_0 $. Nevertheless,  we  checked (not shown here) that   this particular choice does not affect the generality of our results, provided  classical chaos is strong enough to guarantee the ergodicity of the motion of the single spins on 3D-unit spheres. 

The second part of the total Hamiltonian,
\begin{equation}
 V =  J_0
\sum_{k=1}^{L-1}   \sum_{i=k+1}^{L} \frac{1}{|i-k|^\nu}  S_i^x S_{k}^x ,
\label{eq:ham0b}
\end{equation}
describes the spins interaction. They are subjected to a two-body interaction $V$ of strength $J_0$ and  a variable interaction range determined by $\nu$. We set   $J_0 > B_0$, which guarantees strong chaos~\cite{Benet2023} both in the quantum and classical descriptions. 
In what follows, we mostly consider $\nu = 1.4$, which corresponds to short-range interaction and is also referred to as ``weak long-range'' interaction~\cite{Defenu2023}. Additional results for different ranges, $\nu >1$,  are provided in Sec.~\ref{SecNameIV} and show that in the short-range regime, our  results are independent of $\nu$.
 
\subsection{Quantum Model}
The spins are quantized with an integer value $S$ and the effective Planck constant is $\hbar =1/\sqrt{S(S+1)} $, so that the semiclassical limit is achieved for $S \gg 1$. The ``non-interacting many-body basis", in which $H_0$ is a diagonal matrix, corresponds to the eigenstates of $H_0$ and is denoted by 
$ \ket{k} \equiv \ket{s_1,...,s_j,...,s_L}$, where $-S\leq s_j \leq S$ and $j=1,...,L$. The interaction $V$ couples basis vectors that differ by two excitations, so there are two symmetry sectors, each of dimension  $\text{dim} = (2S+1)^L/2$.

{\it Quantum initial state}: The quantum dynamics starts after a quench from $H_0$ to $H$, so that the initial state $|\Psi(0) \rangle $ is a many-body basis vector $\ket{k_0}$
eigenstate of $H_0$.  The components of the evolving wave function at time $t$, written in the many-body noninteracting basis, are 
\begin{equation}
    \label{eq:compq}
    \begin{array}{ll}
     \langle k |\Psi(t)\rangle  =  
  \sum_\alpha  C_k^\alpha \left( C_{k_0}^\alpha \right)^*    e^{-iE_\alpha t/\hbar} ,
    \end{array}
\end{equation}
where 
\begin{equation}
C_k^\alpha \equiv \langle k |\alpha \rangle
\label{eq:cka}
\end{equation}
 and $\ket{\alpha}$  is an eigenstate of the total Hamiltonian $H$ with energy $E_\alpha$.
We consider initial states with energy in the middle of the spectrum, $E_0 = \langle \Psi(0)|H|\Psi(0)\rangle \simeq 0$, where the system has been found to be maximally chaotic~\cite{Benet2023}.

\subsection{Classical Model}
\label{classicalInitial}
The starting point for the  classical model are the classical equations of motion, 
\begin{eqnarray}
\dot{S}_k^x &= & - B_k  S_k^y , \nonumber
\\
\dot{S}_k^y &=&  B_k  S_k^x + J_0 S_k^z \sum_{i\ne k}  \frac{S_i^x}{|i-k|^\nu}, 
\label{eq:eqm}
\\
\dot{S}_k^z &=&  - J_0 S_k^y \sum_{i\ne k}   \frac{S_i^x}{|i-k|^\nu} \nonumber
\end{eqnarray} 
which automatically guarantee the conservation of the angular momentum $|\vec{S_k}|^2=1 $ for each $k=1,...,L$.

The motion of each spin occurs on a 3D unit sphere, as explained in~\cite{Benet2023}. However, the motions of $S^x$ and $S^y$ are principally different from that of $S^z$. If the interaction $J_0$ is very weak, the $k$-th spin rotates around the $z$-axis with  frequency $B_k$, keeping the $S^z$-component almost constant. In contrast, if the interaction is strong, one expects the full coverage of the unit sphere. An important question is then how the chaotic properties of the motion of individual spins emerge with the increase of the spin-spin interaction. The detailed analysis performed in~\cite{Benet2023} revealed the following. 

 From Eqs.(\ref{eq:eqm}) we can get a second-order differential equation for the $S^z$ component of the $k$-th spin,
\begin{equation}
\label{eq:parametric}
\ddot{S_z^k} + \Omega_k^2(t) S_z^k = F_k(t),
\end{equation}
where both 
\begin{eqnarray}
\label{eq:omdrive}
&\Omega_k^2 (t) = J_0^2 \left[\sum\limits_{j\ne k} 
\dfrac{ S_j^x(t)}{|j-k|^\nu} \right]^2 , \\
&F_k (t) = J_0 \sum\limits_{j\ne k} \dfrac{ B_j S_j^y(t) S_k^y(t)- B_k S_j^x(t) S_k^x(t) } {|j-k|^\nu} 
\end{eqnarray}
are quasi-periodic functions defined by the $x$ and $y$ spin components of all other spins $j \neq k$. This means that Eq.~(\ref{eq:parametric}) describes the motion of a {\it parametric oscillator} with a time-dependent frequency $\Omega_k(t)$, under an external quasi-periodic force originated from the motion of the spins $j \neq k$. The important point is that both $F_k(t)$ and $\Omega_k(t)$,  in the first order of perturbation in $J_0$ do not depend on the $S^z_k$-component. This means that the main mechanism of chaos in the considered spin model arises from the parametric instability of the linear oscillator, which is caused by the time-dependence of the frequency $\Omega_k(t)$. In contrast, the instability resulting from the overlap of non-linear resonances only emerges in the second order of perturbation theory, and therefore plays a minor role.

Note that the instability of the motion of the individual spins can be  measured numerically in a relatively straightforward manner, in contrast to the more challenging task of determining the Lyapunov spectra.  Our numerical analysis in~\cite{Benet2023} revealed that the maximal Lyapunov exponent, $\lambda_{+}$, associated with the motion of a single spin is approximately equal to the maximal Lyapunov exponent in the full spectrum of exponents of the many-body spin model. This finding significantly simplifies the derivation of
the characteristic timescale defined by the Lyapunov spectra.   

Since local instability determined by the Lyapunov exponent is not enough to characterize global chaos, that is, chaos on the level of the global phase space, which is typically associated with ergodicity,
we carefully analyzed the problem of classical ergodicity in Ref.~\cite{Benet2023}. 
As explained there,
a very efficient and simple way to rigorously define classical ergodicity in spin systems is to verify the ergodicity of the motion of each individual spin on its unit sphere.  
This significantly simplifies the numerical analysis of 
ergodicity, because we do not need to consider the full many-dimensional phase space. 
 
Specifically, in~\cite{Benet2023}, we proved that the motion of each single spin is ergodic by  (i) finding the  distribution of each one of the three Cartesian components of each single spin, $S_x(t)$, $S_y(t)$, and $S_z(t)$, and (ii) showing that these distributions  follow the expression for each component of the random eigenstates of $3$D random matrices. In this way, we numerically verified that our model  is completely ergodic 
for $J_0 \gtrsim 3 $. This value marks the crossover from a partially chaotic to an ergodic system with strong chaos both in the quantum and classical description. By a partially  chaotic, we mean a system characterized by a positive Lyapunov exponent, but where chaos is not enough to guarantee the full coverage of the available space space.

 {\em Classical initial conditions:} To keep the quantum-classical description as close as possible, we choose a set of classical initial conditions where $S_k^z(0)$ is such that $H_0(0) = \sum_k B_k S_k^z(0) =0 $ and $S_k^{x,y}(0)$ are randomly chosen in $[-1,1]$. This is the classical analogue to the quantum initial state where the values for the spins in the $z$-direction are fixed, so  $\langle S^x_k \rangle = 
\langle S^y_k \rangle =0$, and $E_0 \simeq  0$.

\begin{figure*}[th]
    \centering
    \includegraphics[width=15cm]{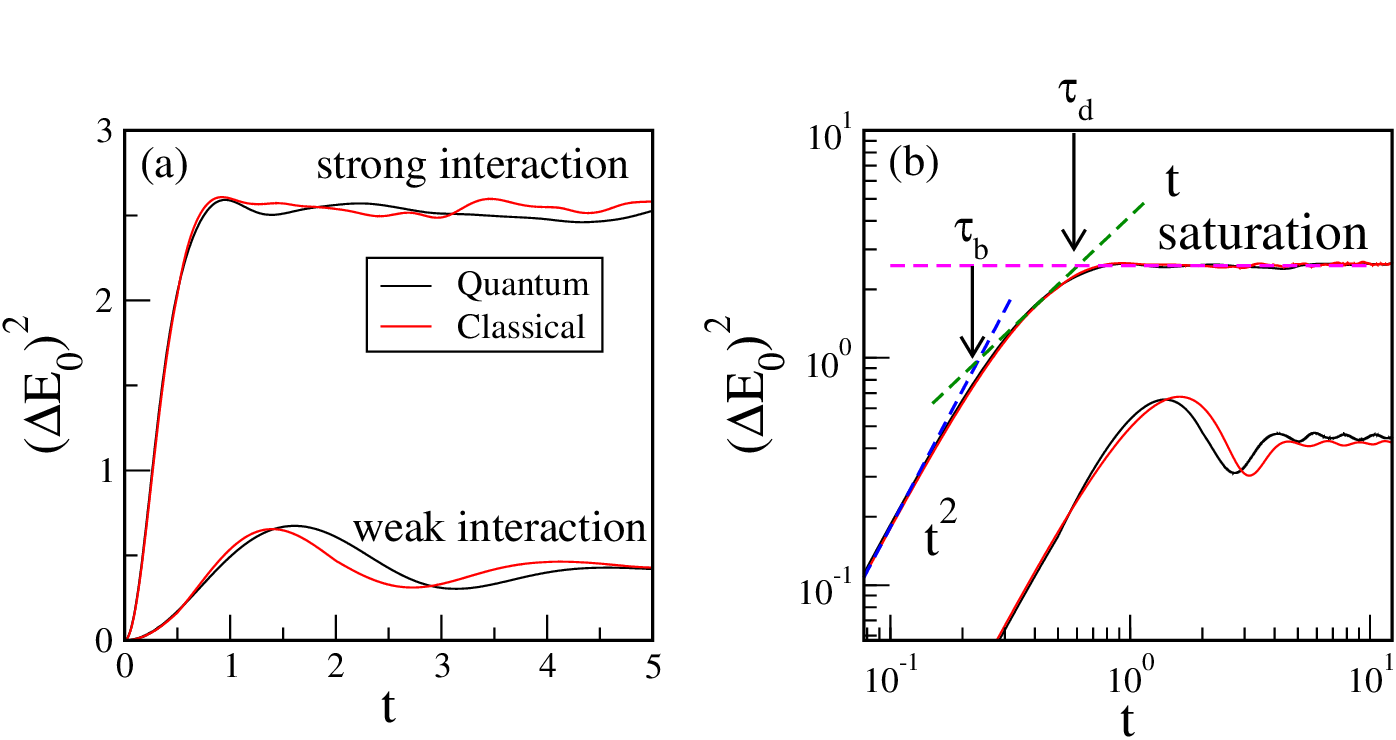}
    \caption{(a) Classical and quantum energy spread $(\Delta E_0)^2(t)$ in time for two interaction strengths:  strong  interaction $J_0=3$ (upper curves)  and weak interaction $J_0=0.6$ (lower curves). Red curves correspond to  classical data and black curves, to quantum data.
    (b) Same as (a) but in the log-log scale to better show the different behaviors at different timescales. Dashed blue line indicates ballistic behavior $(\Delta E_0)^2 \propto t^2$, dashed green lines indicates diffusive behavior $(\Delta E_0)^2 \propto t$, and the saturation of the dynamics, $(\Delta E_0)^2 \propto const$, is indicated with a horizontal magenta line. The vertical arrows indicate the approximate values of $\tau_b$, where the behavior changes from ballistic to diffusive, and $\tau_d$, where the diffusive dynamics saturates.  The parameters are: $L=9$, $B_0=1$, $\delta W=0.2$, $\nu=1.4$.  For the classical case, the average is done over $10^4/L$ initial conditions with  $|E_0|<0.01$. For the quantum case, the average is done over $50$ initial basis states with energy $|E_0|<0.01$. For the quantum simulation, we use the spin quantum number $S=1$.
}
    \label{fig:initial}
\end{figure*} 

\section{Relaxation in the Energy Shell: Global Observable} 
\label{SecEnergy}

The goal of this paper is to determine, both semi-analytically and numerically, the  timescales that characterize the quantum evolution toward equilibrium after a quench, and to explore how these timescales  depend on the parameters of our spin model. To achieve this, we begin by comparing the quantum and classical dynamics of a global observable in this section, followed by the analysis of a local observable in the next section. The analysis of the evolution of a quantity without a classical limit is presented in Sec.~\ref{Sec:NPC}.

In both the quantum and  classical models, the dynamics occur within the energy shell, which is defined by the projection of $H$ onto $H_0$ \cite{Benet2023}. The width of this shell is restricted by the strength of inter-particle interaction, rather than the whole energy space. 

 The global observable that we consider in this section  is the variance in the energies of the single particles given by $H_0$. This quantity spreads in the energy shell due to the inter-particle interaction. In the quantum model, it is defined by the following relation
\begin{equation}
\label{eq:que1}
(\Delta E_0)^2 (t)  = \langle \Psi (t) | H_0^2 |\Psi(t) \rangle -
\langle \Psi (t) | H_0 |\Psi(t) \rangle^2 .
\end{equation}
The corresponding classical quantity is obtained by substituting the quantum average $\langle...\rangle$ with the average over many initial conditions, indicated as $ \overline{(\ldots)}$  (see Sec.~\ref{classicalInitial}).

Notice that in the quantum case, the energy of the initial state, which is an eigenstate of $H_0$, can be equivalently computed in terms of the total Hamiltonian  or of the noninteracting Hamiltonian,
\begin{equation}
 E_0 =   \langle \Psi(0)| H |\Psi(0) \rangle = \langle \Psi(0)| H_0 |\Psi(0) \rangle ,
\end{equation}
because the two-body interaction between spins has zero diagonal matrix elements in the basis of $H_0$. 

To identify the different timescales emerging during the dynamical process, we first compare in Fig.~\ref{fig:initial} the classical and quantum results for $(\Delta E_0)^2 (t) $ numerically obtained for two interaction strengths, weak ($J_0=0.6$) and strong ($J_0=3$) interaction.
The spin quantum number considered is $S=1$. One sees that the correspondence between the quantum and classical results is extremely good even for such small spin number. 

\begin{figure*}[th]
    \centering
    \includegraphics[width=16cm]{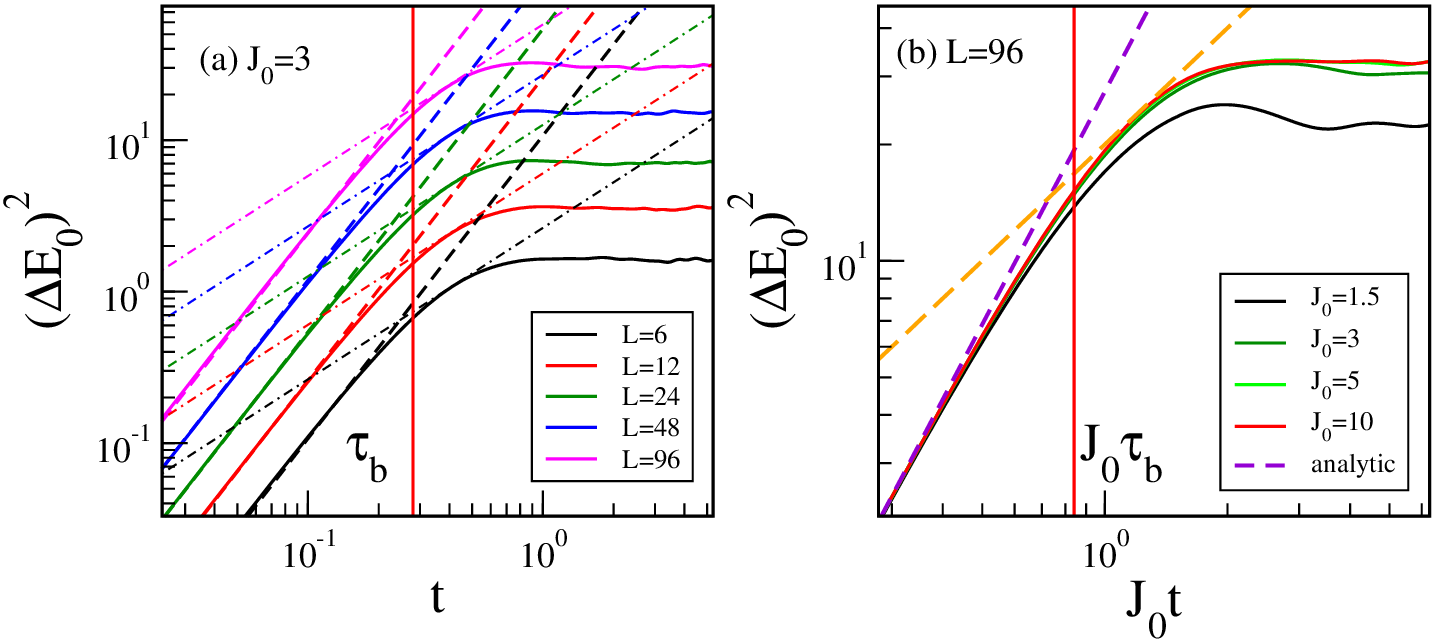}
    \caption{(a) Classical energy spread  $(\Delta E_0)^2 (t)$ in time for a strong  interaction strength, $J_0=3$, and different number of spins $L$, as indicated in the legend. Dashed and dot-dashed lines indicate, respectively, the ballistic and diffusive behaviors. The dashed vertical line marks the time $\tau_b$ in Eq.~(\ref{eq:taub}) at which the dynamics switches from ballistic to diffusive, thus demonstrating the independence on $L$ for this timescale. 
   (b) Classical energy spread as a function of $J_0 t$ for $L=96$ and different values of $J_0$. The curves collapse on each other and the intersection between the ballistic and diffusive regimes is a single point $J_0 \tau_b = const$, thus indicating that $\tau_b \propto 1/J_0$. The purple dashed line corresponds to Eq.~(\ref{eq:bal1}), obtained for the ballistic behavior at short times in the limit of large system size.
   The orange dashed line indicates the diffusive behavior and is obtained using Eq.~(\ref{eq:d1}).  
       The vertical solid line corresponds to Eq.~(\ref{eq:taub}). The average was performed over $10^4/L$ initial conditions with  $|E_0|<0.01$.    The other parameters are 
   $B_0=1, \delta W=0.2, \nu=1.4$.
}
    \label{fig:bal}
\end{figure*} 

In the log-log scale of Fig.~\ref{fig:initial}(b), it becomes evident that the curve for strong interaction ($J_0=3$) exhibits three different dynamical regimes. The dynamics is initially ballistic (blue dashed line), then it becomes diffusive (green dashed line), before finally relaxing to equilibrium (magenta horizontal dashed line). On the other hand, the diffusive regime is absent for  weak interaction ($J_0=0.6$), so saturation happens after the ballistic spread. These results are consistent with the findings in Ref.~\cite{Benet2023}, where it was numerically proved that for strong interaction, the motion is not only chaotic (defined by a maximal positive Lyapunov exponent), but  also ergodic on the unit sphere of each spin. In contrast, for weak interaction, where diffusion is absent, the dynamics is not ergodic, even though the presence of a maximal positive Lyapunov exponent signals the presence of classical chaos. Despite these differences, we observe in Fig.~\ref{fig:initial}, that the QCC holds independently of the interaction strength and over all timescales.

\subsection{Ballistic regime}
In this subsection, we perform a semi-analytical study of the shortest timescale, which is characterized by ballistic propagation. To do this, we turn to the classical model to derive analytical estimates, which are then compared  with numerical data.

For short time, the variance  $(\Delta E_0)^2(t)$ is proportional to  $t^2$.  To find the time $\tau_b$ at which the spreading of energy switches from ballistic to diffusive, we first need to find the velocity $v_b$ defined by the equation  $\Delta E_0 = v_b t$. This in turn  can be obtained from the classical equations of motion by expanding $S_k^z(t)$ for short time, 
$$S_k^z(t) = S_k^z(0) +t \dot{S}_k^z (0) +(1/2) t^2 \ddot{S}_k^z(0) + O(t^3) $$ 
and taking into account that we choose initial conditions to have $E_0(0) = 0$.  As mentioned before,  this means that the $z$-component  of all spins, $S_k^z(0)$, initially leads to $\sum_k B_k S_k^z (0)= 0$  and the $x$ and $y$ components are chosen to be completely random (keeping fixed the unit length for the spin vector). 

Taking the ensemble average over the initial random conditions and using Eq.~(\ref{eq:eqm}), we can show that
\begin{eqnarray}
\label{eq:bal}
(\Delta E_0)^2 &=&  t^2 \overline{ \left( \sum_{k=1}^L  B_k J_0 \sum_{j\ne k}\frac{S_k^y(0) S_j^x(0)}{|k-j|^{2\nu}}\right)^2 } \!\!\!\! +O(t^3) \nonumber \\
&=&  \displaystyle \frac{t^2}{9}  \left( \sum_{k=1}^L  B_k^2 J_0^2 \sum_{j\ne k} \frac{1}{|k-j|^{2\nu}} \right) + O(t^3) 
\end{eqnarray}
where the last equality is due to our choice of completely random  $x$ and $y$ components, so that the only non-zero terms are $\overline{ S_k^y(0)^2} = \overline{S_j^x(0)^2}= 1/3$. The analytical expression in Eq.~(\ref{eq:bal}) is plotted in Fig.~\ref{fig:bal}(a) (dashed lines) and compared with numerical results (full curves) for different system sizes $L$ and a strong interaction strength that guarantees the ergodic motion. We reiterate that the dashed lines characterizing the short time dynamics are not fitting lines, but Eq.~(\ref{eq:bal}).

\begin{figure*}[t]
    \centering
    \includegraphics[width=16cm]{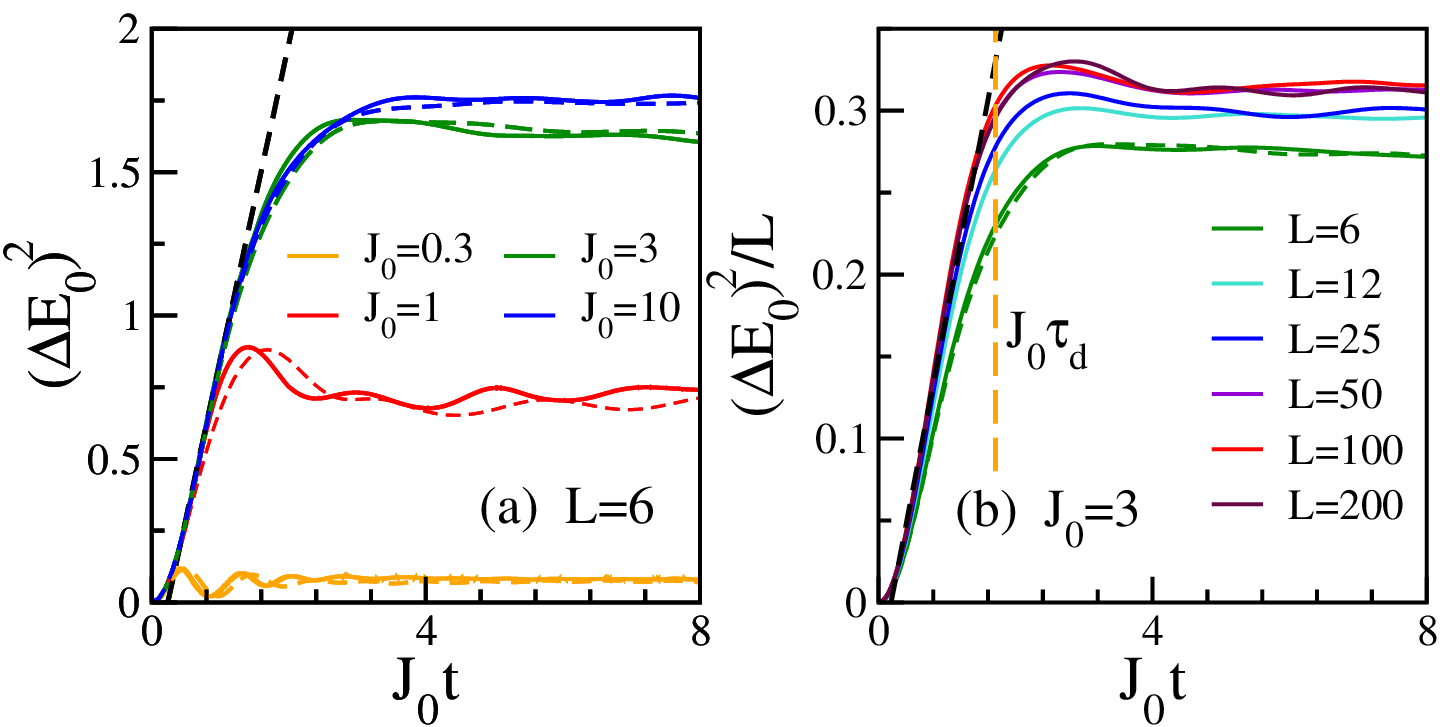}
    \caption{(a) Energy spread   $(\Delta E_0)^2(t)$ as a function of the renormalized time $J_0 t$ for a fixed number of spins, $L=6$,  and different interaction strengths $J_0$, and (b) renormalized energy spread {\it vs}  $J_0 t$ for  a fixed  interaction strength,  $J_0=3$, and different system sizes $L$. Quantum (dashed lines) and classical (solid lines) data are shown. The black dashed line in both panels indicates the linear (diffusive) growth with slope given by the fitted diffusion coefficient $D$ in Eq.~(\ref{eq:d}) for $J_0=10$ in (a) and $L=200$ in (b).  The vertical dashed line in (b) marks the diffusion timescale $\tau_d$. For the classical case, we average over $10^4/L$ initial conditions with  $|E_0|<0.01$. For the quantum case, we average over $50$ initial basis states with energy $|E_0|<0.01$, and $S=2$. 
}
    \label{fig:diff}
\end{figure*} 

In the limit of large system size, Eq.~(\ref{eq:bal}) can be further simplified as
\begin{equation}
\label{eq:bal1}
\begin{array}{lll}
  &(\Delta E_0)^2 (t) \equiv v_b^2 t^2  = \left( \sum\limits_{k=1}^L  B_k^2 J_0^2\sum\limits_{j\ne k} 
  \frac{1}{|k-j|^{2\nu}} \right) \displaystyle\frac{t^2}{9}  \\ 
  &\\
  &\simeq   \displaystyle \frac{2}{9} \sum\limits_{k=1}^L  B_k^2  J_0^2 \sum\limits_{j=1}^\infty \dfrac{1}{j^{2\nu}} t^2 \  \simeq  \dfrac{2}{9} L \langle B^2\rangle J_0^2 \zeta(2\nu)  t^2,
\end{array}   
   \end{equation}
where in the last equality, we defined implicitly the Riemann zeta function $\zeta(2\nu)$, which is finite for $\nu > 1/2$, and the second moment of the single-particle frequencies 
\begin{equation}
\label{eq:b2}
\langle B^2 \rangle = \frac{1}{L} \sum_{k=1}^L B_k^2.
\end{equation}
The ballistic velocity is therefore
\begin{equation}
\label{eq:vel}
v_b = J_0 \sqrt{ \frac{2L\langle B^2 \rangle \zeta(2\nu) }{9}} . 
\end{equation}
For our particular choice for the single-particle frequencies, 
$B_k = B_0 +\delta B_k$, where $\delta B_k$ is a small random shift in the interval $(-\delta W, \delta W)$, we can further simplify Eq.~(\ref{eq:b2}) as  
\begin{equation}
\label{eq:dw}
\langle B^2 \rangle \simeq B_0^2 + \frac{\delta W^2}{3}.
\end{equation}

We test Eq.~(\ref{eq:bal1})  in Fig.~\ref{fig:bal}(b) for a sufficiently large system size ($L=96$) and different interaction strengths. The collapse of the curves, indicate good agreement with that equation. 
In addition to strong interactions ($J_0\geq 3$), we also include an example of moderate interaction strength ($J_0=1.5$), 
where the energy shell is not completely filled  (compare the black and red curves) and thus ergodicity is not fully achieved.

To summarize,  the ballistic motion is described by the following relation,
\begin{equation}
\label{eq:bal2}
(\Delta E_0)^2 (t) = v_b^2 t^2 \qquad  {\rm with } \qquad v_b = v_0 J_0  \sqrt{L},
\end{equation}
where we stress the dependence of $v_b$ on the system size  $L$  and the interaction strength $J_0$. The constant
\begin{equation}
    \label{eq:v0}
v_0 = \frac{1}{3}\sqrt{2 \zeta(2\nu) \langle B^2\rangle}
\end{equation}
depends only on the interaction range $\nu$ and on the second moment $\langle B^2\rangle$ of the single-particle frequencies. For our choices of parameters, $v_0 \simeq 0.53$. 

In what follows, we use $v_b$ to find the ballistic time $\tau_b$ at which the ballistic regime ends and the diffusion process starts, provided that chaos is strong. To estimate $\tau_b$, we also need the analytical dependence of the diffusion coefficient $D$ on the model parameters, which is the subject of the next subsection. It is interesting to see how the dependence of $v_b$ and $D$ on $L$ combine to guarantee that $\tau_b$ is independent of the system size. Furthermore, as we will see in Sec.~\ref{Sec:taud}, the diffusion time $\tau_d$ also turns out to be independent of $L$.

\subsection{Diffusive regime}

As seen in Fig.~\ref{fig:initial}(b), the variance of the single-particles energies after $t \approx \tau_b$ grows linearly in time when the interaction is strong, which allows us to write a ``diffusion-like" relation,
\begin{equation}
\label{eq:d1}
(\Delta E_0)^2 (t) \simeq  D t,
\end{equation}
and associate $D$ with a diffusion coefficient. In Fig.~\ref{fig:diff}, we use quantum number $S=2$ and show  $(\Delta E_0)^2(t)$ for a fixed system size $L$ varying the interaction strength $J_0$ [Fig.~\ref{fig:diff}(a)]  and for a fixed strong interaction
$J_0$ varying  the system size $L$ [Fig.~\ref{fig:diff}(b)]. 

In Fig.~\ref{fig:diff}(a), the system size is relatively small ($L=6$) to make possible the comparison with the quantum dynamics. 
We deduce from this figure that the slope of the linear growth is proportional to the interaction strength, so $D \propto J_0$. By rescaling the variance to the system size, we observe in Fig.~\ref{fig:diff}(b) that the curves for large values of $L$ are superimposed. This indicates that for large system sizes, we also have $D \propto L$, while for small $L$, finite-size effects are relevant.  Combining these results we get the equation
\begin{equation}
\label{eq:d}
D  = c_0  J_0 L,
\end{equation}
where $c_0\approx 0.2$ is a constant obtained with a linear fitting. We stress that the diffusion-like spreading in the energy shell is independent of the choice of parameters, provided they ensure strong quantum chaos. 
The underlying mechanism of this diffusion-like dynamics may have a similar origin to that of the celebrated  kicked rotator model, which is a 1D time-dependent system~\cite{Casati1979}. In this model, quantum diffusion, characterized by the linear increase in time of the second moment of the energy, follows closely the classical diffusion up to a certain time. Nevertheless, while classical diffusion is irreversible~\cite{Chirikov1981} due to local exponential instability associated with the classical dynamics,  quantum diffusion is reversible, due to the linearity of Schr\"odinger equation.
It is an open question whether a  picture similar to the one developed for the kicked rotator could be extended to our  many-dimensional system.

Equating Eq.~(\ref{eq:bal2}) and Eq.~(\ref{eq:d1}), 
$$(\Delta E_0)^2 = v_b^2 \tau_b^2 = D \tau_b,$$ 
and using Eq.~(\ref{eq:d}),
we can get an estimate for the time $\tau_b$ at which the diffusion starts, 
\begin{equation}
\label{eq:taub}
 \tau_b = \frac{c_0} {J_0 v_0^2} = \frac{9 c_0}{2\zeta^2(2\nu)} \frac{1}{J_0 \langle B^2 \rangle},
\end{equation}
where the latter equality is obtained by substituting the value of $v_0$  given in Eq.~(\ref{eq:v0}).
This estimate indicates that the time at which diffusion starts is independent of $L$ and  is  inversely proportional to  the interaction strength $J_0$. While it is understandable that the diffusion process  should start earlier if one increases the inter-particle interaction, the independence of $\tau_b$ on the system size $L$ might seem unexpected at a first sight. As we show in the next section, this also occurs for local observables characterized by a well-defined classical limit.

Our results are numerically confirmed in Fig.~\ref{fig:bal}. In  Fig.~\ref{fig:bal}(a), where the interaction strength is  large and different values of $L$ are considered, we mark  the intersection between the lines that give the ballistic  and the diffusive behaviors. As indicated with a  vertical solid line, these crossing points and therefore $\tau_b$ are independent of the system size $L$. On the other hand, in Fig.~\ref{fig:bal}(b), where a large system size ($L=96$) and different interaction strengths are considered, one sees that  $\tau_b$ depends on $J_0$. In this panel, the energy spreading is shown as a function of the renormalized time $J_0 t$. The fact that all curves collapse into a single one, so that one can draw a single line for the ballistic behavior and a single line for the diffusive behavior, indicates that $J_0 \tau_b$ is a constant, so $\tau_b \propto 1/J_0$. The analytical expression for $\tau_b$ in Eq.~(\ref{eq:taub}) is shown in Figs.~\ref{fig:bal}(a)-(b) as a vertical red line.

To estimate at which time the diffusion stops and equilibration sets in, we first need to estimate the saturation value. In the next subsection, we find an approximate expression for the saturation value as a function of the interaction strength $J_0$.

\subsection{Relaxation to the steady state}
\label{Sec:taud}

In Ref.~\cite{Benet2023} we showed that for a sufficiently large interaction strength, $J_0 \gtrsim  3$, and a sufficiently large number of spins, $L > 50$, the classical motion of each single spin in the unit sphere is ergodic. We now use this result to compute the maximal energy spreading in the energy shell due to ergodicity.

Under complete ergodicity, $S_k^z$ can be thought of as  a random independent variable with   $\overline{S_k^z(t)} = 0$ and $\overline{S_k^z(t)^2} = 1/3$. Using this result in the definition of the energy for non-interacting spins,
 \begin{equation}
 E_0 (t) = \sum_{k=1}^L B_k S_k^z(t),    
 \label{eqsm:e0}
 \end{equation}
we obtain the maximal classical energy spreading,
\begin{equation}
(\Delta E_0)^2_{\text{erg}}  = \overline{E_0^2(t)} - \overline{E_0(t)}^2  =  \sum_k B_k^2 \overline{ S_k^z(t)^2} = \frac{1}{3} L \langle B^2 \rangle \\
\label{eqsm:de0}
\end{equation}
where we set  $\overline{S_k^z(t)S_j^z(t)} = 0$ for $k\ne j$. We can further approximate this expression using Eq.~(\ref{eq:dw}),
 \begin{equation}
\label{eqsm:der}
 (\Delta E_0)^2_{\text{erg}}  = \frac{L}{3}  \left(B_0^2 + \frac{ \delta W^2}{3} \right)  .
\end{equation}
This is the energy spreading for completely random variables, as in the case of fully ergodic motion. Inserting our parameters $B_0=1$ and $\delta W=0.2$, we obtain that $$ \Delta E_{rms} \equiv \sqrt{(\Delta E_0)^2_{\text{erg} } }\simeq 0.58 \sqrt{L}.$$  Notice that the width of the ergodic spreading of energy, which is $\propto \sqrt{L}$, is much smaller than  the range of possible values obtained for $E_0(t)$, which is $[-\sum_k B_k, \sum_k B_k]$ and thus proportional to $L $.

\begin{figure}[h!]
    \centering
    \includegraphics[width=7cm]{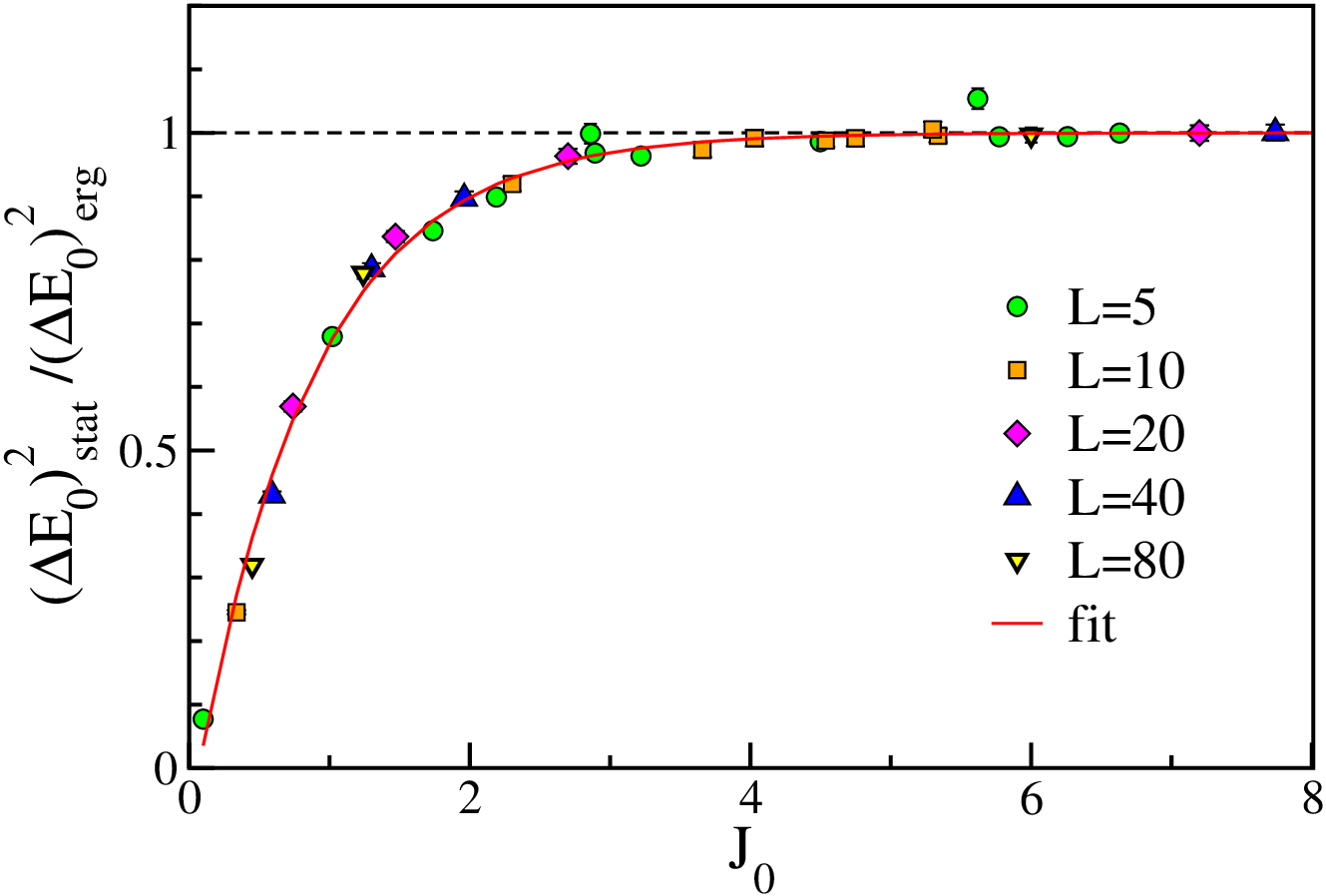}
    \caption{Stationary classical energy spreading (symbols) as a function of the interaction $J_0$ for different system sizes $L$ compared with the ergodic energy spreading in the energy shell (horizontal dashed line). The horizontal dashed line stands for the ergodic spreading $(\Delta E_0)^2_{stat} = (\Delta E_0)^2_{\text{erg}}$, as given in Eq.~(\ref{eqsm:der}), while the red curve is the fitting with the function $f(x)=1-a e^{-bx}$, according to Eq.~(\ref{eq:DeltaEFit}), where $a=1.09$ and $b=1.19$ are the best fitting parameters.  
}
    \label{fig:max}
\end{figure}

In Fig.~\ref{fig:max}, we compare the numerical results obtained for the stationary value 
\begin{equation}
    \label{eq:stat}
    (\Delta E_0)^2_{stat} = \lim_{T\to \infty} \frac{1}{T} \int_0^T \ dt \
    (\Delta E_0(t))^2
    \end{equation}
with the analytical result in Eq.~(\ref{eqsm:de0})  as a function of $J_0$. The saturation value of the energy spreading agrees with the analytical calculation for the ergodic spin motion when $J_0 \gtrsim 3$, thus confirming once more the ergodicity of the motion for strong interaction. For smaller values of the interaction strength, when the motion is not fully ergodic, we achieve an approximate expression for the saturating value of the energy spreading by fitting our data with a two-parameters function, 
$f(J_0) = 1-a e^{-bJ_0}$, which gives
\begin{equation}
    \label{eq:DeltaEFit}
     (\Delta E_0)^2_{stat} =  ( 1-a e^{-bJ_0} )   (\Delta E_0)^2_{\text{erg}} .
\end{equation}
 The above equation is accurate for all $J_0$ values, as shown with the red curve in Fig.~\ref{fig:max}. 
The fitting function is compatible with the fact that there is no energy spreading in the absence of interaction ($J_0 = 0$).

We now have the necessary ingredients to  estimate the relaxation time, $\tau_d$, for the energy spreading using the relation 
\begin{equation}
D \tau_d = (\Delta E_0)^2_{stat}.  
\end{equation}
In the case of the fully ergodic motion, we can substitute $(\Delta E_0)^2_{stat}$ with the analytical expression in Eq.~(\ref{eqsm:der}), $(\Delta E_0)^2_{\text{erg}} $, leading to 
\begin{equation}
\label{eq:taug}
\tau_d =   \frac{ \sum_{k=1}^L B_k^2}{c_0 L J_0} = \frac{\langle B^2\rangle}{3 c_0 J_0},
\end{equation}
This estimate shows that the relaxation time for the energy spreading is independent of the system size, which is numerically confirmed  in Fig.~\ref{fig:diff}(b).

\begin{figure}[h!]
    \centering
    \includegraphics[width=7cm]{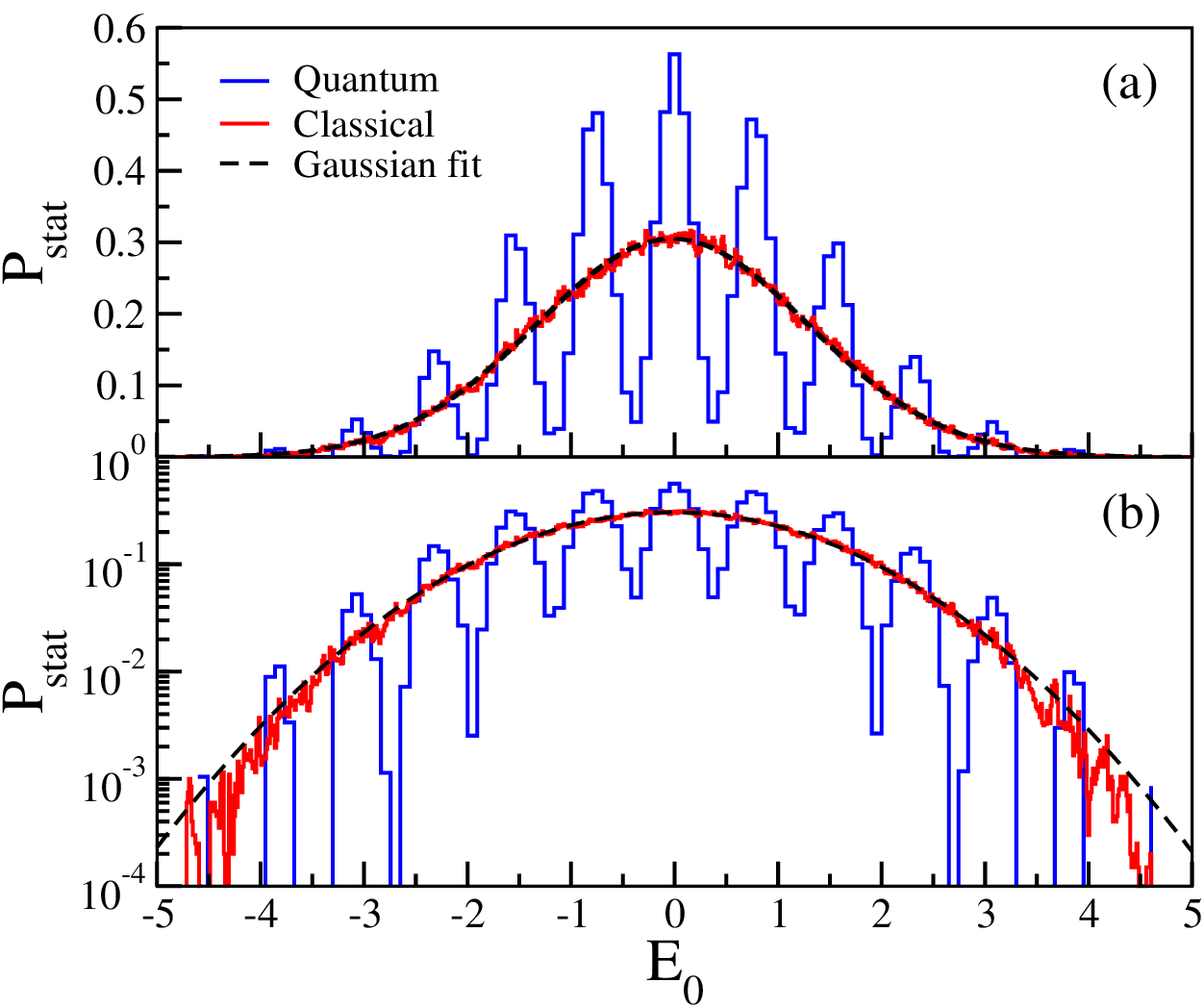}
    \caption{Comparison between classical and quantum stationary distributions of the non-interaction energy for $L=6, J_0=3$. The initial states are the same as those reported in  Fig.1. Spin quantum number $S=2$.  Panel (a) is in normal scale, while (b) is in semi-log scale to show the Gaussian tails.}
    \label{fig:stat}
\end{figure}

Due to the finite  size of the energy shell, the variance $(\Delta E_0)^2(t)$ for time $t \gg \tau_d$ saturates. For a real diffusive process, we expect the stationary energy distribution to become Gaussian.  This is indeed what happens for the classical model, as seen in Fig.~\ref{fig:stat}, with the exception of the far tails [Fig.~\ref{fig:stat}(b)] . 
The quantum distribution, on the other hand, exhibits a clear band structure enveloped by a Gaussian distribution. The absence of a Gaussian shape prevents the association with ``true'' diffusion, even though $(\Delta E_0)^2$ spreads linearly in the quantum domain. Nevertheless, since there are $2LS+1$ bands and the total size of the energy shell is $\sim 2L$, the quantum distribution approaches the classical one for   $S\gg 1 $ at fixed $L$.

To summarize the results of this section so far, we have found that there are two timescales, one at which the diffusive behavior starts and the other where it ends.  Both times are proportional to $1/J_0$, which is physically understandable, {\it and both are independent of the number $L$ of spins.} In the next subsection, we compare the diffusion time with the Lyapunov timescale.

\subsection{Local instability: The Lyapunov  timescale}

Since the single-particle energy spreading occurs in a diffusive-like manner, one could expect it to be governed by local instabilities of the motion. Instability is the main mechanism for diffusion, because it is associated with random trajectories and chaos.

Motivated by the role of the Lyapunov timescale, $\tau_{\lambda}$, in chaotic systems~\cite{Steinberg2019,Malishava2022,Malishava2022PRL,Correale2023,Wang2021b,Bilitewski2021}, we investigate whether this timescale plays any role in the description of the relaxation to equilibrium of $(\Delta E_0)^2(t)$.  Despite the link between chaos and diffusion, these two timescales, $\tau_{\lambda}$ and $\tau_d$, {\it do not need to be necessarily equal}.

\begin{figure}[!h]
    \centering
    \includegraphics[width=7cm]{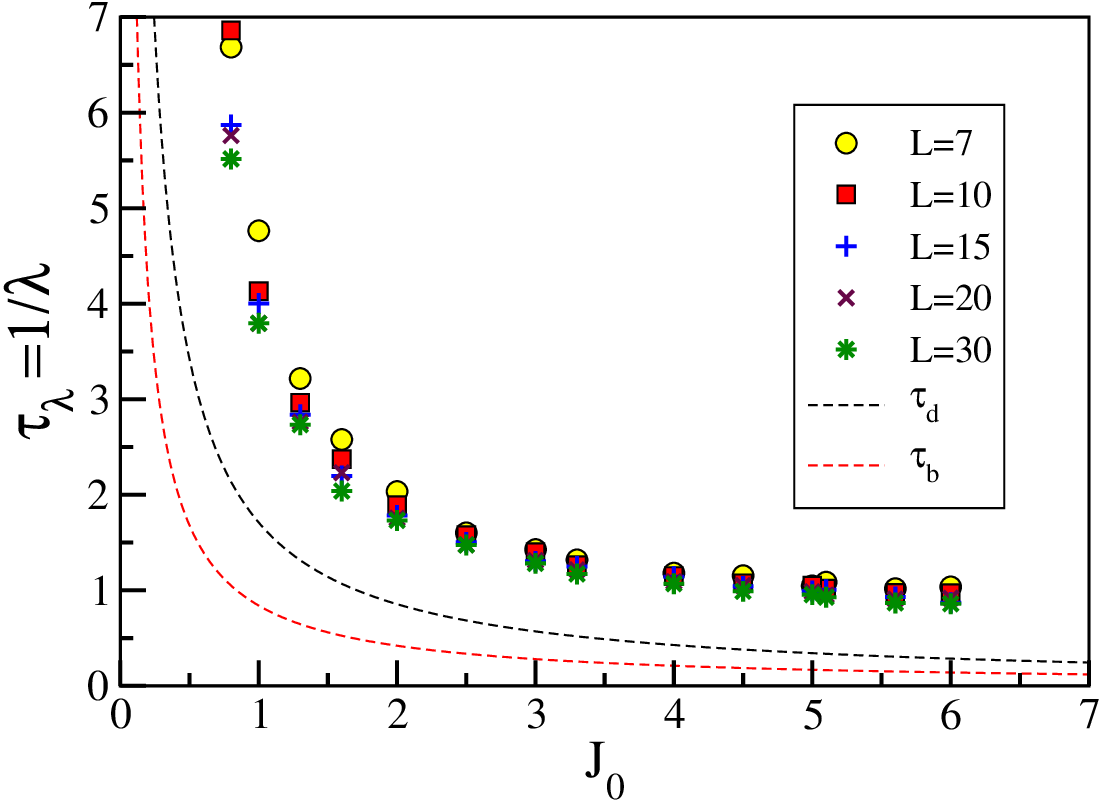}
    \caption{Comparison between the inverse of the maximal Lyapunov exponent (Lyapunov time) for various system sizes $L$ with the timescales for ballistic, $\tau_b$ [Eq.~(\ref{eq:taub})], and diffusive, $\tau_d$ [Eq.~(\ref{eq:taug})], energy spreading as a function of the interaction strength $J_0$.    }
    \label{fig:lyap}
\end{figure}

\begin{figure*}[t]
    \centering
    \includegraphics[width=15cm]{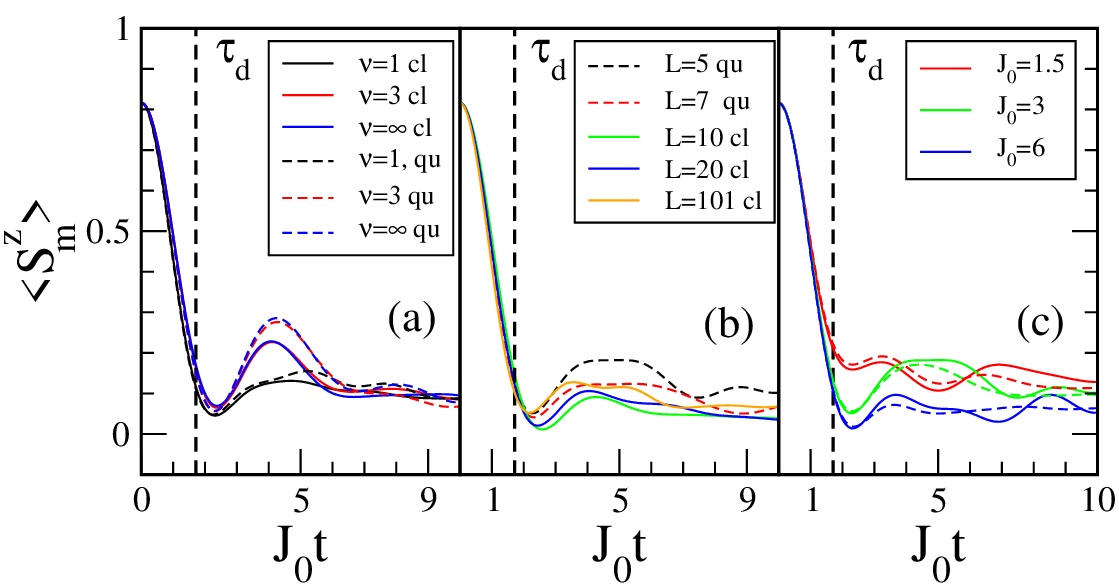}
    \caption{Quantum (dashed curves) and classical (solid curves) evolution of the $z$-component of the central spin for: (a) different ranges $\nu$ of the interaction with fixed system size $L=5$ and $J_0=3$;  (b) different system sizes $L$ with fixed $\nu=1.4$ and $J_0=3$;  (c)  different interaction strengths $J_0$ with $L=5$ and $\nu=1.4$. The vertical lines stand for the diffusion timescale $\tau_d$ obtained  in Eq.~(\ref{eq:taug}). The quantum initial state is $\ket{\Psi (0)} =|0,...,0,S,0,...,0\rangle$.  The classical initial condition corresponds to $S_{m=\lfloor L/2 \rfloor}^z(0) = S/\sqrt{S(S+1)}$ and $S_k^z(0) = 0$ for $k\ne m=\lfloor L/2 \rfloor$, and the $x,y$-components are chosen at random, apart from the $m-th$ spin for which $S_{m}^x=S_{m}^y=0$. We use $10^4/L$ classical initial conditions. For all quantum data    $S=2$.  
}
    \label{fig:local}
\end{figure*}

Even though the many-body system is characterized by the spectrum of all Lyapunov exponents (see \cite{Benet2023}), the maximal exponent sets the smallest time scale for instability. For this reason, we study the maximal  Lyapunov $\lambda_{\text{max}}$ exponent averaged over many different initial conditions with the same single-particle energy $E_0 = 0$. We stress that even though the maximal Lyapunov exponent decreases as the interaction strength decreases, it remains nonzero for weak interaction, $J_0 <1$. It is only at the integrable limit, $J_0=0$, that $\lambda_{max}=0$.
For weak interaction, there is absence of classical  ergodicity, but the model is still chaotic~\cite{Benet2023}.

In Fig.~\ref{fig:lyap} we plot the Lyapunov time, which is the inverse Lyapunov exponent, $\tau_\lambda = 1/\lambda_{max}$, as a function of the interaction strength $J_0$ for different system sizes $L$. We see that $\tau_\lambda$ is slowly dependent of $L$, and  for large interaction strengths, $J_0 \gtrsim 3 $, it is nearly independent of $L$.  The figure suggests that the behavior of the $\tau_\lambda$ with $J_0$ is comparable to that of $\tau_d$ in the range of interactions  where diffusion is observed.

\section{Relaxation of single-spin magnetization: Local observable} 
\label{SecNameIV}

Having established the timescales for the energy spread in the energy shell, we now move our attention to a local observable, namely, to the $S_z$ component of an individual spin. The initial many-body state that we choose for the quantum model is 
$$\ket{\Psi (0)} =|0,...,0,S,0,...,0\rangle,$$ 
where only the spin $m= \lfloor L/2 \rfloor$ in the middle of the chain has maximal value $S$ along the $z$-direction, while all other $L-1$ spins have zero value for the $z$-component. 
The corresponding classical initial condition is  $ S_{\lfloor L/2 \rfloor}^z (0) = S/\sqrt{S(S+1)}$ for the central spin, $S_{k}^z (0) = 0  $ for the other spins, and the $x,y$ components are randomly chosen keeping the length of the spins fixed.  We investigate the time that it takes for the excitation on site  $\lfloor L/2 \rfloor$ to get shared with the other $L-1$ spins and whether the characteristic time for the relaxation is the same $\tau_d$ as obtained 
in Eq.~(\ref{eq:taug}).

We show the evolution of the onsite magnetization of the central spin $m$ for different ranges $\nu$ of the interaction in Fig.~\ref{fig:local}(a),  for different system sizes in Fig.~\ref{fig:local}(b), and for different interaction strengths in Fig.~\ref{fig:local}(c). The results in Fig.~\ref{fig:local}(a) indicate that when the interaction is short range,  the relaxation time does not depend on the value of $\nu$. The excellent QCC in all panels justifies the use of the classical dynamics for the analysis of large system sizes performed in Fig.~\ref{fig:local}(b). This plot makes it evident that the timescale for the relaxation is independent of $L$. 
Figure~\ref{fig:local}(c) demonstrates that the relaxation time depends on the interaction strength, similar to the dependence observed for the energy spreading.

\begin{figure*}[t]
   \centering{}
    \includegraphics[width=15cm]{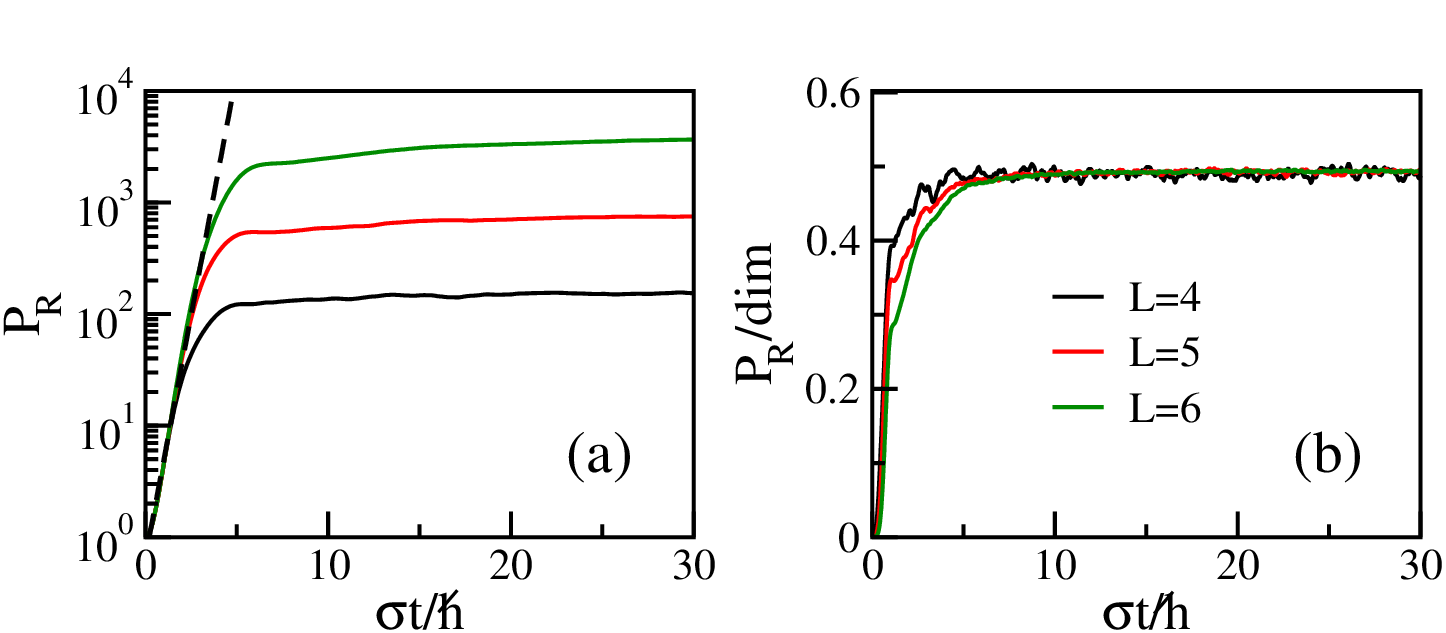}
    \caption{(a) Participation ratio, $P_{R}$, and (b) participation ratio  renormalized by the dimension of the Hilbert space as a function of time  renormalized by the width of the LDoS, $\sigma t/\hbar$,  for different system sizes $L$. The dashed line in (a) is $e^{2\sigma t/\hbar}$. The average is performed over 50 initial  states in the energy range $|E|<0.01$, $J_0=3$, $S=2$. 
    }
    \label{fig:9npc}
\end{figure*}

To compare the relaxation time for the onsite magnetization with that for the energy spreading, we mark with a vertical dashed line in Figs.~\ref{fig:local}(a)-(c), the  diffusion time $\tau_d$ obtained 
in Eq.~(\ref{eq:taug}) and find good numerical agreement with it.
Therefore, our results presented in Fig.~\ref{fig:diff} and Fig.~\ref{fig:local}  confirm  that both the local and the global quantities considered here exhibit the same relaxation timescale, independently of the length $L$. 

To explain why the relaxation times for the magnetization and energy spread are the same and coincide for both the classical and quantum cases, independently of the system size $L$, we turn to the classical model. We focus on the evolution of $S_k^z(t)$, as described by Eq.~(\ref{eq:parametric}), where the time-dependent frequency 
$\Omega_k^2 (t)$ and the driving nonlinear force $F_k (t)$ are
given by Eq.~(\ref{eq:omdrive}) and depend on all other spin components $j\neq k$. As discussed in Sec.~\ref{classicalInitial}, these equations reveal that the main mechanism of chaos in this system is linear, that is, chaos arising from the parametric instability of the linear oscillator. Taking this property into account, it becomes clear that increasing the system size  $L$ simply adds more harmonics to the expressions for the driving force $F_k$ and  time-dependent frequency $\Omega_k(t)$. Since even a small number of incommensurate frequencies $B_k$ is enough to produce  effective randomness, adding more frequencies does not significantly alter the results. This explains why increasing $L$ has a minimal effect on the dynamics of individual spins in their motion on the unit sphere.

\section{Quantum Observable without a classical limit}
\label{Sec:NPC}

In the previous sections, we investigated quantities that had a classical limit. We now analyze a quantum observable  that has no a classical analogue, namely the participation ratio,
\begin{equation}
\label{eq:npc}
P_{R}(t)= \frac{1}{\sum_k |\langle k | \Psi(t)\rangle |^4} .
\end{equation}
This quantity  is purely quantum, because it measures the effective number of many-body basis states 
 $| k\rangle $ occupied  by the evolved state $|\Psi(t) \rangle $ at  time $t$. Changing the basis representation changes the value of $P_{R}(t)$, so no classical limit can be defined. This quantity describes global relaxation in the Hilbert space of quantum states.

Knowledge of the energy distribution of the initial state, the so-called LDoS, helps with the description of the evolution of  $P_{R}(t)$. The LDoS is defined as
\begin{equation}
    \label{eq:LDOS}
    W_{k_0}(E) = \sum_\alpha \delta(E-E_\alpha) |C_{k_0}^\alpha|^2 , 
\end{equation}
where the coefficients $C_{k_0}^\alpha = \langle \alpha |k_0\rangle$, as in Eq.~(\ref{eq:cka}), and $|k_0\rangle$ is an initial state corresponding to a non-interacting basis state, typically taken in the middle of the energy spectrum. The width of the LDoS is given by 
\begin{equation}
    \sigma= \sum_{k \neq k_0}\langle k|H |k_0 \rangle = \sum_{\alpha} |C_{k_0}^\alpha|^2 E^2 - \left(\sum_{\alpha} |C_{k_0}^\alpha|^2 E \right)^2 .
    \label{Eq:sigmaLDOS}
\end{equation}

When the initial state is composed of many chaotic eigenstates of the total Hamiltonian, $P_{R}(t)$ grows exponentially in time with a rate given by the width of the LDoS~\cite{Borgonovi2019b}, as shown, for instance, in Fig.~\ref{fig:9npc}(a). To extract a reliable estimate for the relaxation timescale, in Fig.~\ref{fig:9npc}(b), we rescale $P_{R}(t)$ to the dimension of the subspace associated with the initial state and verify that all curves saturate at the same point. For values of $J_0$ that ensure quantum chaos, the saturation point of $P_{R}(t)$ is roughly $\text{dim}/2=(2S+1)^L/4$.
With this result, we can find an analytical estimate of the timescale $\tau_N$ for the relaxation of $P_{R}(t)$ using the equality
\begin{equation}
 \label{eq:npc}  
 P_{R} (\tau_{N}) \simeq e^{2\sigma \tau_{N} /\hbar } =  (2S+1)^L/4,
\end{equation}
which gives 
\begin{equation}
\label{eq:tn1}
\tau_{N} \propto L \hbar \ln (2S+1) / \sigma     .
\end{equation} 
In the equation above, $\sigma$ is the width of the quantum LDoS written in Eq.~(\ref{Eq:sigmaLDOS})

\begin{figure}[h]
    \centering
    \includegraphics[width=7cm]{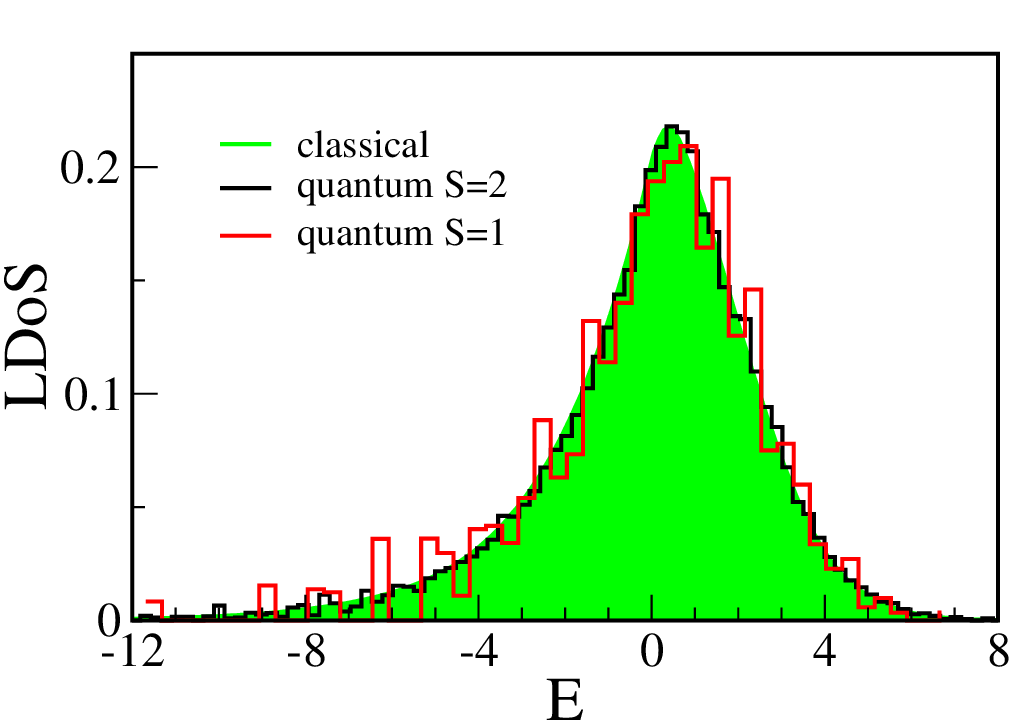}
    \caption{Quantum and classical LDoS for $L=6$, $J_0=3$. The histograms for the quantum case are for spin number $S=1$ (red) and $S=2$ (black). The shaded histogram represents the classical LDoS.
      }
    \label{fig:ldos}
\end{figure}

Obtaining analytical estimates for the width of the quantum LDoS is challenging, as it requires the exact diagonalization of Hamiltonian matrices that grow exponentially with $L$. Nevertheless, using the quantum-classical correspondence exploited in Ref.~\cite{Benet2023}, it is possible to numerically build the classical LDoS. The agreement between the classical and quantum LDoS in the quantum chaotic regime is remarkable even for small spin quantum numbers $S=1,2$ as shown in Fig.~\ref{fig:ldos}. This allows us to use the classical LDoS to estimate the width $\sigma_{cl}$ for large system sizes. This is an important result, because it implies that instead of the diagonalization of huge matrices, we can extract information about the LDoS by simply integrating  $3L$  differential equations, which is feasible for systems as large as $L=10^2$ spins with a standard laptop. This ``semi-quantal'' approach was also discussed in~\cite{Borgonovi2002}.

Using the classical equations of motion, we arrive at
\begin{eqnarray}
    \sigma_{cl}^2 &= &  \frac{J_0^2}{9} \sum_{k=1}^L\sum_{j>k} \frac{1}{|j-k|^{2\nu}} \equiv \frac{J_0^2}{9} \zeta(\nu, L),
    \label{eqsm:var}
\end{eqnarray}
where  the symbol ``$\equiv$''  defines implicitly the function 
$\zeta(\nu, L)$ for any $\nu$ and finite $L$.
For large values of $L$, this function can be approximated as 
\begin{equation}
    \label{eqsm:zeta}
 \zeta(\nu,L) \simeq (L-1)\sum_{k=1}^\infty \frac{1}{k^{2\nu}} = (L-1)\zeta(2\nu),  
\end{equation}
where $\zeta(2\nu)$ is now the Riemann zeta function.
Thus, for sufficiently large $L$, we get that 
\begin{equation}
    \label{eq:zeta1}
  \sigma_{cl} \simeq \frac{J_0\sqrt{L-1} \zeta(2\nu)}{3}.  
\end{equation}
This result holds for any $\nu > 1/2$, when the Riemann zeta function converges. 

\begin{figure}[t]
    \centering
    \includegraphics[width=7cm]{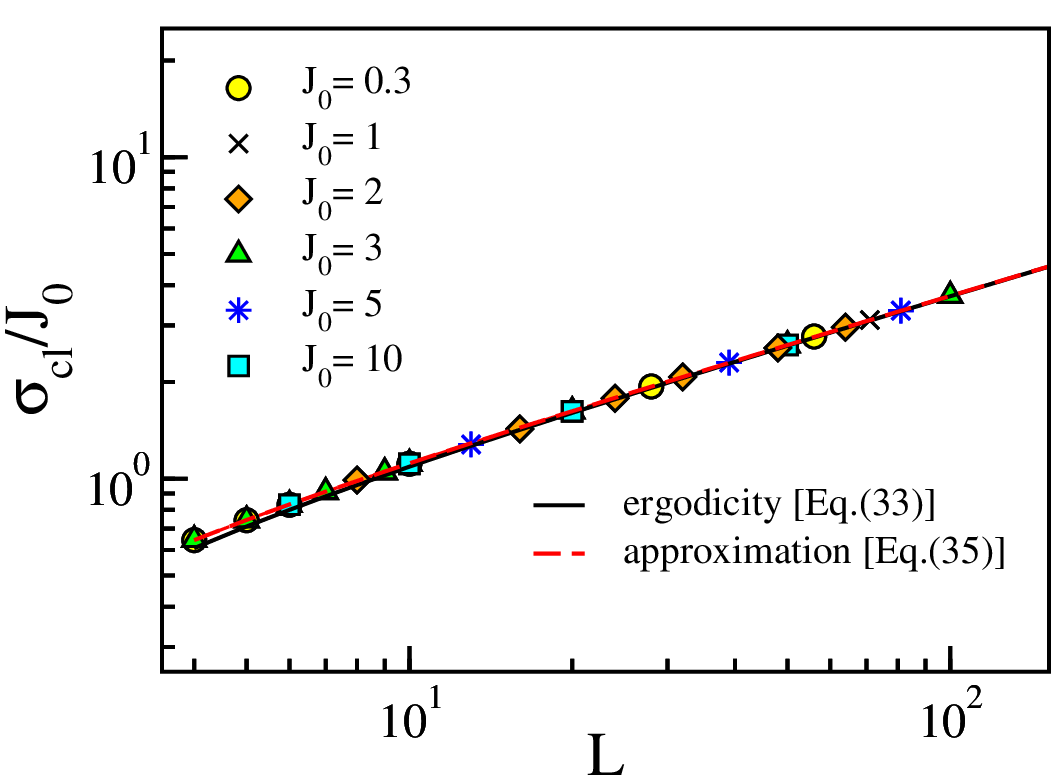}
    \caption{Classical width of the LDoS divided by the interaction strength, $\sigma_{cl}/J_0$, as a function of the system size $L$. Full (black) line is the ergodic approximation for finite $L$ in Eq.~(\ref{eqsm:var}) and the red dashed curve stands for the  expression for the ergodic approximation for large $L$ in Eq.~(\ref{eq:zeta1}). 
      }
    \label{fig:cldos}
\end{figure}

In Fig.~\ref{fig:cldos}, we compare numerical results for the classical width of LDoS with the exact expression for finite $L$ in Eq.~(\ref{eqsm:var}) and the approximate expression in Eq.~(\ref{eq:zeta1}). As one can see, even for $L \gtrsim  10$, the approximate expression matches extremely well the numerical data.

Using the analytical result for the classical width of LDoS in Eq.~(\ref{eq:zeta1}), we finally get the estimate for the relaxation time for $P_{R}(t)$,
\begin{equation}
   \label{eq:taun}
    \tau_{_N} \propto \sqrt{L} \frac{1}{J_0} \frac{\ln(2S+1)}{\sqrt{S(S+1)}}.
\end{equation}
The above equation shows that for any fixed $S$, the relaxation time increases with $L$, as indeed noticeable in Fig.~\ref{fig:9npc}(a). 
This indicates that the quantum relaxation process is more complex than the classical one. Furthermore, comparing the relaxation time for the energy spread and onsite magnetization with  that for the participation ratio, we infer  that different quantum observables may relax on distinct timescales.

\section{Conclusion}
\label{Sec:Discussions}

We investigated the quantum-classical correspondence (QCC) of many-body spin systems to analyze their relaxation dynamics following a quench. In the regime of strong chaos, we verified that the quantum and classical dynamics are analogous even for small spin values such as $S=1,2$. The QCC allows for the use of the classical system to estimate the timescales of very large quantum systems. This approach enabled us to derive semi-analytical results for the timescales governing the spread of the single-particles energies, $(\Delta E_0)^2(t)$, and the relaxation of the $z$-magnetization of individual spins, $\langle S_k^z(t) \rangle$.

The analysis of $(\Delta E_0)^2(t)$  revealed three distinct temporal regimes. The first one, arising from perturbation theory, corresponds to the ballistic spread of energy up to $\tau_b$. Subsequently,  the energy spread exhibits a diffusive-like behavior that persists until $\tau_d$, when the dynamics saturates due to the finite width of the energy shell. Supported by the results in Ref.~\cite{Benet2023}, where it was shown that for strong interaction the classical motion of each spin is ergodic in its unit sphere,  we confirmed that both the ballistic and diffusive timescales remain independent of the system size $L$ (at least for sufficiently large $L > 10$).  

The fact that diffusion is conditioned to the existence of chaos prompted the question of how the Lyapunov timescale, $\tau_{\lambda}$, defined as the inverse of the Lyapunov exponent, compares with the diffusion time $\tau_d$. The Lyapunov time characterizes the local instability of the motion of individual spins, while the diffusion time determines the global energy spread in the energy shell. In principle, there is no reason for these two timescales to coincide. However,  our numerical results revealed that $\tau_{\lambda}$ and $\tau_d$ are of the same order and, once again, do not depend on the system size.

Our analysis of the classical and quantum evolution of  $\langle S_k^z(t) \rangle$ demonstrates that the relaxation time for this local quantity, like that for the global quantity $(\Delta E_0)^2(t)$, does not dependent on $L$ either. We expect this result to be general and confirmed for other physical observables with a well-defined classical limit.

We found that the diffusion time depends on the single-particle frequencies $B_k$ and the interaction strength $J_0$, following the relation $\tau_d \propto \langle B^2 \rangle/J_0$. While one might expect similar results   when the frequencies are nearly constant or completely random, the expression for $\tau_d$ naturally raises the question of whether the frequencies could be engineered to induce a system-size dependence in the relaxation process. Such an approach could offer a tool for controlling the dynamical properties of the system. 

A closer examination of the classical equations of motion provided a more detailed justification for why the relaxation time is independent of the system size. Our interpretation can be summarized as follows.
The second-order differential equation for $S_k^z$ describes a parametric oscillator. The   time-dependent frequency and force of the oscillator are not significantly affected by an increase in system size.  

Motivated by previous studies~\cite{Borgonovi2016,Borgonovi2019,Borgonovi2019b,Lezama2021} on the relaxation process of the participation ratio, $P_R(t)$, this quantity was included in this paper,  despite its lack of a classical limit.
As demonstrated analytically and verified numerically, in the region of strong quantum chaos, $P_R(t)$ grows exponentially  before reaching saturation. The exponential growth is governed by the width of the Local Density of States (LDoS). As explained in~\cite{Benet2023}, this width can be obtained from the classical counterpart of the model, which allows us to access large system sizes. We find that the relaxation of the participation ratio  is proportional to $\sqrt{L}$. This implies that, unlike observables with a well defined classical limit, the participation ratio thermalizes  on a timescale that increases with system size. This result highlights the importance of the chosen observable in determining the timescales for thermalization. It may also provide an explanation for the various results for the relaxation time  reported in the literature, where different $L$-dependencies have been observed for quantities such as survival probability and participation ratio.

\begin{acknowledgements}
F.B.  acknowledges support by the Iniziativa Specifica INFN-DynSysMath  and  MIUR within the Project No. PRIN 20172H2SC4. F.M.I. acknowledges financial support from CONACyT (Grant No. 286633). L.F.S. was supported by  NSF Grant No. DMR-1936006. 
This research was supported in part by grants NSF PHY-1748958 and PHY-2309135 to the Kavli Institute for Theoretical Physics (KITP).    L.F.S and F.M.I.  gratefully acknowledge support from the Simons Center for Geometry and Physics, Stony Brook University at which some of the research for this paper was performed.
\end{acknowledgements}

\bibliography{biblioNSF2019}

\end{document}